# A harmonized and interoperable format for storing and processing polysomnography data

Riku Huttunen[1,2], Matias Rusanen[1,2], Sami Nikkonen[1,2], Henri Korkalainen[1,2], and Samu Kainulainen[1,2]

[1]Department of Technical Physics, University of Eastern Finland, Kuopio, Finland

[2]Diagnostic Imaging Center, Kuopio University Hospital, Kuopio, Finland

Date of current version: 20 November 2023

## Corresponding author
Riku Huttunen

Email address: riku.huttunen@uef.fi

Postal address: Yliopistonrinne 3, P.O. BOX 1627 (Canthia), FI-70211, Kuopio, Finland

Co-author email addresses: matias.rusanen@uef.fi, sami.nikkonen@uef.fi, henri.korkalainen@uef.fi, samu.kainulainen@uef.fi

## Funding statement
This study was financially supported by the Research Committee of the Kuopio University Hospital Catchment Area for the State Research Funding (projects 5041789, 5041803, 5041804, 5041807, and 5041809), by the Emil Aaltonen Foundation, by the Finnish Cultural Foundation (Central Fund and Kainuu Regional Fund), by the Regional Council of Pohjois-Savo (ERDF project A77427), by Tampere Tuberculosis Foundation, and by the Research Foundation of the Pulmonary Diseases.

## Conflicts of interest
The Authors declare that there are no conflicts of interest.

## Abstract

Polysomnography (PSG) data is recorded and stored in various formats depending on the recording software. Although the PSG data can usually be exported to open formats, such as the European Data Format (EDF), they are limited in data types, validation, and readability. Moreover, the exported data is not harmonized, which means different datasets need customized preprocessing to conduct research on multiple datasets. In this work, we designed and implemented an open format for storage and processing of PSG data, called the Sleeplab format (SLF), which is both human and machine-readable, and has built-in validation of both data types and structures. SLF provides tools for reading, writing, and compression of the PSG datasets. In addition, SLF promotes harmonization of data from different sources, which reduces the amount of work needed to apply the same analytics pipelines to different datasets. SLF is interoperable as it utilizes the file system and commonly used file formats to store the data. The goal of developing SLF was to enable fast exploration and experimentation on PSG data, and to streamline the workflow of building analytics and machine learning applications that combine PSG data from multiple sources. The performance of SLF was tested with two open datasets of different formats (EDF and HDF5). SLF is fully open source and available at https://github.com/UEF-SmartSleepLab/sleeplab-format.

## Introduction

Various file formats are used in different polysomnography (PSG) recording software. These file formats can usually only be accessed using the commercial software they operate with. Luckily most software support exporting the recordings to European Data Format (EDF) [1] or its extension EDF+ [2]. However, the EDF files also need specialized software for reading and exploration of the data. All signals in the EDF files need to be stored as 16-bit integers, and the storage of metadata is limited. In addition, EDF does not provide validation or compression of the data. Moreover, due to the usage of custom and insufficiently tested software that write the files, EDF files often contain non-standard data and format defects. This leads to incompatibility problems, such as errors when the files are read. Without consistent tools for reading and writing PSG data both in files and in memory, a substantial part of research resources is used in implementing or customizing software for new use cases.

As standardizing PSG software-specific proprietary data formats is unfeasible, one solution is to deal with the heterogeneous formats as the first step of the data processing pipeline. This step only needs to be conducted once; when the originally heterogeneous datasets are stored in a harmonized format, downstream processing is simplified. Ideally, running the same processing pipeline (e.g., performing automated analyses or training a machine learning model) for different datasets, as well as for combinations of datasets, should not require any changes to the pipeline.

In this work, we designed a data format for PSG recordings, called the Sleeplab format (SLF). The file format is both human and machine-readable and utilizes the underlying file system for hierarchical data structures. In addition, we developed a software that can be used to read, write, and validate the data conforming to the format. This reader-writer loop ensures the correctness of data structures and types, enabling efficient reuse of code and results, as well as flexible combination of datasets originating from various sources. The software is openly available at https://github.com/UEF-SmartSleepLab/sleeplab-format.

# Methods

## Sleeplab format

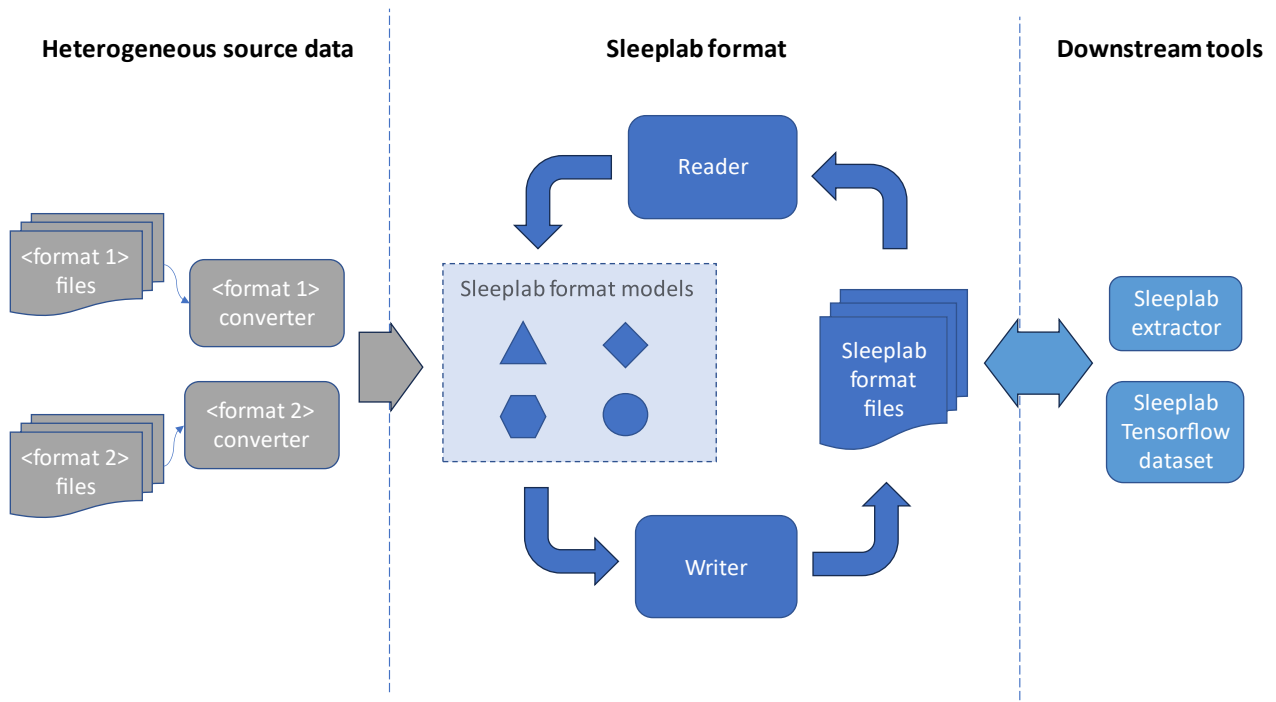

*Figure 1: Overview of the Sleeplab format and its ecosystem. After converting the heterogeneous source data formats into Sleeplab format, the data structures are validated always when reading or writing. This enables implementing downstream tools that work on different datasets without customizations.*

The basic building blocks and the workflow for utilizing SLF are illustrated in Figure 1. The heterogeneous source data represents the various existing PSG recording formats. To convert these formats to SLF, a custom software needs to be implemented. Since the source data formats can be arbitrary, the converters are not part of SLF, although examples are made available. During the conversion, the original dataset is first read to the in-memory SLF models. Then, the dataset is written into the SLF files utilizing the SLF writer. The SLF files in turn can be read back to memory with the SLF reader. The SLF models validate that the data conforms to the format every time the data is read or written.

In addition to the converters and the format itself, SLF ecosystem contains downstream tools that utilize the SLF for storage and processing. One example is the Sleeplab extractor (available at https://github.com/UEF-SmartSleepLab/sleeplab-extractor), which can be used to extract a subset of the SLF dataset, preprocess it, and write the resulting dataset back to disk using the SLF writer. Another example of downstream tools is the Sleeplab Tensorflow Dataset library (available at https://github.com/UEF-SmartSleepLab/sleeplab-tf-dataset), which provides methods to read a SLF dataset as a Tensorflow dataset directly from the SLF files.

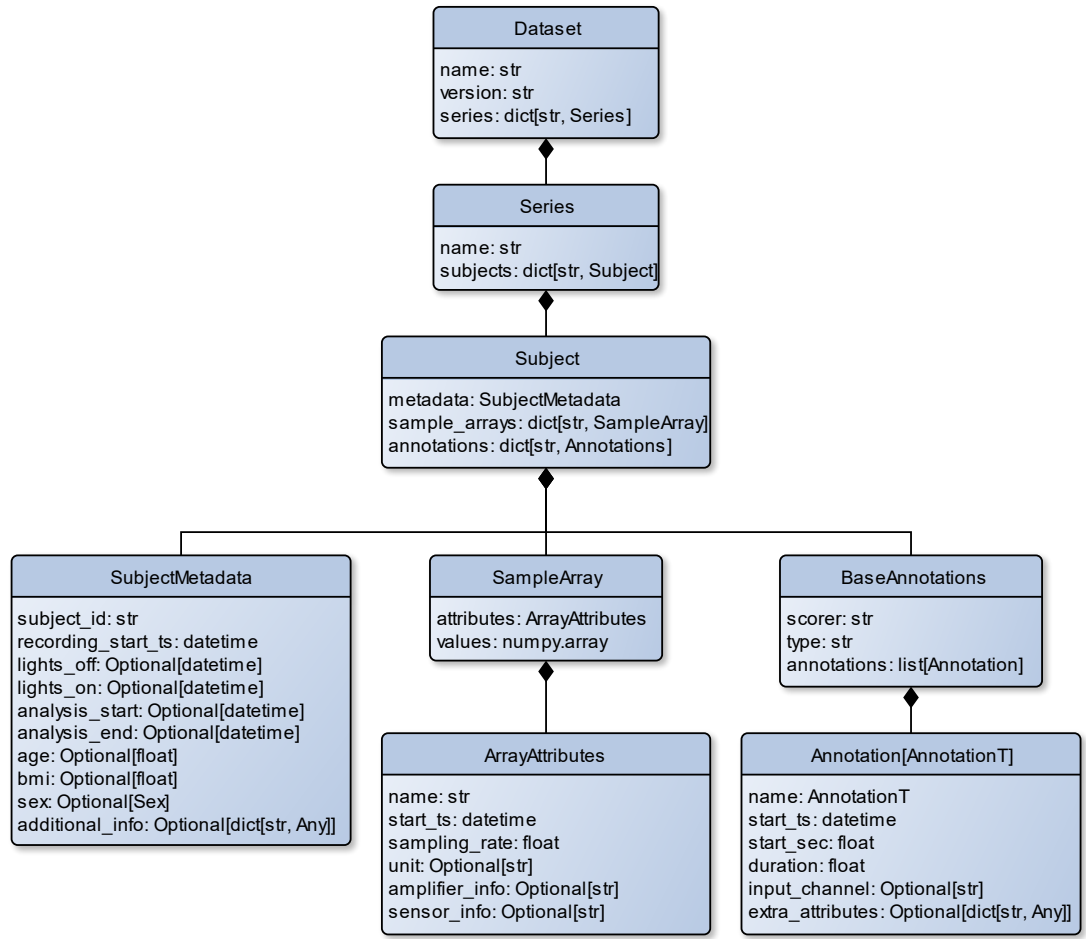

*Figure 2: Diagram of the base classes and their composition in the Sleeplab format. A dataset consists of any number of series that contain subjects. Each subject has metadata, and any number of sample arrays and annotations.*

In the core of SLF are the SLF models, which are classes that specify the entities and datatypes that a SLF dataset can contain. Class diagram of the base models is shown in Figure 2. From top down, a dataset contains any number of series, and a series contains any number of subjects. Each subject has metadata, such as the recording start time and age. The subject also has any number of sample arrays which contain, for example, the signals recorded during the PSG. In addition, the subject can have any number of annotations, which are defined as segments with a start time, duration, and a name. The Annotation class is parameterized with the type of annotation name, which enables restrictions to the naming conventions. For example, a sleep stage annotation may allow wakefulness, stage N1, stage N2, stage N3, and rapid eye movement (REM) sleep as the name, following the current rules and terminology set by the American Academy of Sleep Medicine.

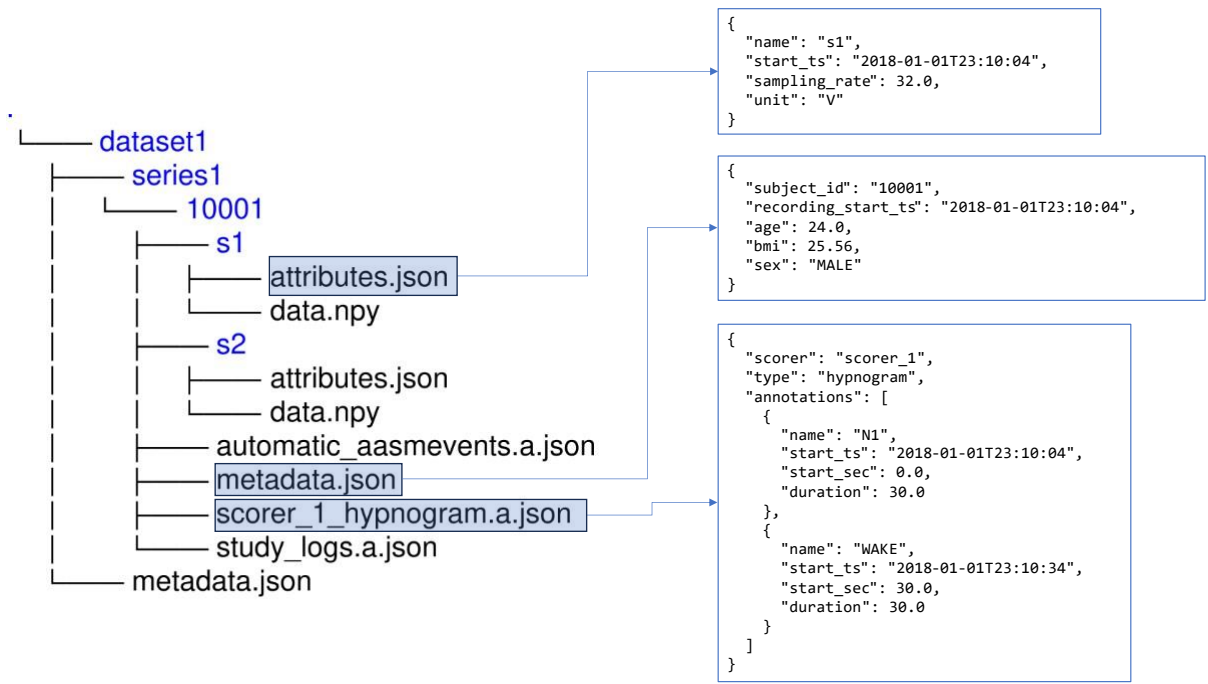

*Figure 3: File and folder structure of a Sleeplab format dataset. The hierarchy of datasets, series, and subjects is reflected in the folder structure. All metadata is stored in plain-text JSON files.*

The SLF writer utilizes the file system for the class hierarchy. An example of the file and folder structure is shown in Figure 3. Each dataset, series, and subject is a folder. All metadata and annotations are stored in plain-text JSON files. This enables using any file manager and text editor to explore a SLF dataset. The JSON files are not compressed as their file size compared to the numerical data is negligible and leaving them uncompressed preserves the human-readability of the format. Only the sample arrays, which contain most of the data, are stored in binary format as storing them in a text-based format such as CSV would harm the read performance and increase the file size considerably. Each sample array is stored in its own folder, containing the binary data and an *attributes.json* file that contains metadata such as the name and sampling rate of a PSG signal.

### Software

The software was developed and tested using Python 3.10 and Ubuntu 22.04. The SLF models were implemented using Pydantic [3], which adds data types and validation on top of Python classes and enables writing the models to JSON files and reading the JSON files back to the models.

The numerical data in SLF, i.e., the sample arrays, can be stored in either uncompressed or compressed format. The uncompressed data is stored in NumPy files [4]. The signals can be stored in any of the NumPy data types. If compression is desired, SLF utilizes Zarr [5], which allows choosing from various compression algorithms and stores the metadata in separate JSON files along with the compressed arrays. For compression, SLF uses Zstandard, which is a lossless compression algorithm that is fast especially when decompressing the data during reading. The compression level of Zstandard, ranging from -7 to 22, can be specified when using the SLF writer.

### Performance assessment

SLF was benchmarked with two openly available datasets: Sleep-Cassette from Sleep-EDF [6], [7] and Dreem open datasets [8]. The Sleep-Cassette dataset consists of 153 full-night PSG recordings with manually scored sleep stages [6]. The Dreem open datasets consist of full-night recordings with manually scored sleep stages for 25 healthy volunteers and 55 obstructive sleep apnea patients [8]. The size on disk, read performance, and write performance were assessed. For write performance, the overall duration of the conversion from source files to SLF files was measured. For read performance, the overall duration to read the dataset and calculate the mean of each sample array was measured. Mean calculation was included to ensure that the data is read into memory. Three different compression scenarios for the sample arrays were tested: uncompressed, Zstandard compression level 9, and Zstandard compression level 22 (the maximum level of Zstandard). The benchmark source code is available at https://github.com/UEF-SmartSleepLab/sleeplab-format/tree/main/examples/benchmark.

The benchmark was run on a server with a Western Digital Black 5TB 7200RPM SATA hard drive, AMD Ryzen Threadripper 2990 WX CPU, and 128 GB of DDR4 memory. Caching files to the memory was disabled to assess performance reading directly from the hard drive. In the performance assessment, PyEDFlib [9] was used to read the EDF files of the Sleep-Cassette dataset. To read the HDF5 files of the Dreem open datasets, the h5py package was used [10].

## Results

*Table 1: A comparison of size and performance between the benchmarked configurations for the Sleep-Cassette (n=153) and Dreem open datasets (n=80).*

| Dataset | Format | Data type | Compression | Size (GB) | Conversion time (s) | Read time (s) (speed (MB/s)) |
|---|---|---|---|---|---|---|
| SC | EDF | int16 | - | 7.60 | - | 161 (47.2) |
| SC | SLF (NumPy) | float32 | - | 15.2 | 548 | 101 (150) |
| SC | SLF (Zarr) | float32 | Zstd, level 9 | 6.02 | 976 | 85.8 (70.2) |
| SC | SLF (Zarr) | float32 | Zstd, level 22 | 5.13 | 5410 | 85.4 (60.0) |
| DOD | HDF5 | float32 | - | 60.4 | - | 200 (-) |
| DOD | SLF (NumPy) | float32 | - | 31.4 | 490 | 204 (153) |
| DOD | SLF (Zarr) | float32 | Zstd, level 9 | 29.0 | 850 | 271 (107) |
| DOD | SLF (Zarr) | float32 | Zstd, level 22 | 29.0 | 4720 | 280 (103) |

SC = Sleep-Cassette; DOD = Dreem open datasets; EDF = European Data Format; HDF = Hierarchical data format; int16 = 16-bit integer; float32 = 32-bit floating-point; Zstd = Zstandard.

The performance assessment results are reported in Table 1. When compressing the sample arrays, the size of the EDF data was remarkably reduced from 7.6 GB down to 5.13 GB when using Zstandard with the maximum compression level 22. With the HDF5 files, the size reduction due to compression was only minor when compared to the uncompressed SLF files (31.4 GB vs. 29.0 GB). The size of the HDF5 files was almost twice the size of the uncompressed SLF files (60.4 GB vs. 31.4 GB) because the original HDF5 files contained large quantities of unaccounted space. Since the unaccounted space was not read into memory, the read speed for the HDF5 files is not reported in Table 1.

With the Sleep-Cassette dataset, the uncompressed SLF files were faster to read (101 s) when compared to the EDF files (161 s). The read times were further decreased when the SLF data was compressed (85.8 s and 85.4 s with compression levels 9 and 22, respectively). With the Dreem open datasets, read times for the uncompressed SLF files and the HDF5 files were similar (204 s vs 200 s). The compressed SLF files were slower to read (271 s and 280 s with compression levels 9 and 22, respectively).

# Discussion

In this work we presented SLF, an open-source data format for harmonization, storage, and processing of PSG data. The main motivation to develop SLF was to streamline the exploration and use of various PSG datasets for research purposes. From the start, SLF was designed to be open, transparent, and interoperable to support utilization of large PSG datasets without the repetitive use of resources to decipher and read the data from different formats to consistent structures. Moreover, once the data is in an open format, there is no need to purchase multiple software licenses to read the proprietary file formats, cutting the costs of research. Harmonization of the data along with tools to reliably read and write the data enables faster experimentation and increases reproducibility when conducting research. In addition, the harmonization allows applying the same analytics pipelines to different datasets without additional data wrangling once the original datasets have been converted to SLF. This is especially useful in machine learning research in which training and testing the algorithms with different combinations of datasets is essential. SLF has already been used to harmonize different datasets in our research [11]. There is ongoing work to harmonize several large datasets for multicenter PSG data research.

Most of the PSG data is in the biosignal measurements that require the highest sampling rates, such as the electroencephalography (EEG) signals. The signal datatypes dictate the size of the uncompressed data. For example, the EDF files store the data as 16-bit integers, while 32-bit floating-point numbers were used with SLF. Thus, the EDF dataset was half of the size of the uncompressed SLF dataset (Table 1). However, the HDF5 dataset was twice the size of the uncompressed SLF dataset, although the signals in the HDF5 files were also 32-bit floating-point numbers. Closer inspection revealed that the original HDF5 files contain large amounts of unaccounted space, which roughly doubles the file size.

Most physiological signals contain noise that cannot be losslessly compressed. Thus, if no information loss has occurred due to data processing or changes in data types, the lossless compression methods cannot significantly reduce the size of these signals. This was the case with the HDF5 dataset, where the size of the uncompressed SLF data was only reduced from 31.4 GB to 29.0 GB when using Zstandard with the maximum compression level (Table 1). However, the size of the EDF dataset was reduced remarkably when using compression (7.6 GB vs. 5.13 GB), even when 32-bit floating-point numbers were used instead of the 16-bit integers used by the EDF files. Moreover, the size of the dataset was further reduced when a higher level of compression was used. We suspect that this reduction of size due to compression is related to information loss due to data type conversions in the process of writing and reading the EDF files, although we did not conduct a systematic study on the underlying reasons.

Whether to use compression is a tradeoff between the size reduction, disk input/output (I/O) performance, and computational overhead of the compression algorithm. With the EDF dataset, the size reduction due to compression was so large that it was faster to read the compressed signals and decompress them than to read the uncompressed signals (Table 1). The disk used in the tests was a SATA-connected HDD with maximum read speed of around 200 MB/s. With slower network storage, the performance gains of compression can be even larger. With the HDF5 dataset, the compressed data was slower to read compared to uncompressed data, and size reduction due to compression was minor. Thus, it is sensible to store the HDF5 dataset as uncompressed SLF files in the tested environment.

SLF utilizes NumPy's memory-mapping when reading data from the uncompressed sample arrays. This means that only the parts of signals that are accessed are read into memory, which makes the SLF reader memory efficient and reduces the amount of disk I/O due to reading. This allows performing fast, memory efficient computations reading parts of the signals directly from disk. For example, an iterative machine learning algorithm trained with smaller segments of the full PSG recordings can read only the data needed for the current training step. Similarly, analyses that use specific annotated events can only read to memory those annotated parts of the signals.

Read performance, memory-efficiency, and keeping the disk I/O low are essential when utilizing data on disk for analytics. One scenario to take advantage of both memory-mapping and compression is to store the full datasets in compressed format, and extract study-specific subsets of possibly downsampled signals in uncompressed format. The smaller uncompressed subset of data can then be utilized in a more performant computational environment with possibly more limited storage space.

Currently, before converting to SLF the data needs to be exported from the PSG software to an intermediate format, such as EDF. The conversion requires customized code, and some of the limitations of the exported format are propagated to the SLF files. However, as the format is open, we hope that in the future the PSG software could support directly exporting to SLF.

## Conclusion

SLF is a both human and machine-readable format to store and process PSG data. The software provides tools for reading and writing that validate that the data is always in the correct format. In addition, SLF enables the use of various data types and compression algorithms for numerical data. SLF focuses on harmonization of PSG data from different sources, which enables efficient combination of datasets and increases code reusability. The software is fully open source and can be easily extended for different use cases.

## References

[1]  B. Kemp, A. Värri, A. C. Rosa, K. D. Nielsen, and J. Gade, "A simple format for exchange of digitized polygraphic recordings," *Electroencephalogr. Clin. Neurophysiol.*, vol. 82, no. 5, pp. 391–393, May 1992, doi: 10.1016/0013-4694(92)90009-7.
[2]  B. Kemp and J. Olivan, "European data format 'plus' (EDF+), an EDF alike standard format for the exchange of physiological data," *Clin. Neurophysiol.*, vol. 114, no. 9, pp. 1755–1761, Sep. 2003, doi: 10.1016/S1388-2457(03)00123-8.
[3]  S. Colvin *et al.*, "pydantic/pydantic: v2.3.0." Zenodo, Aug. 23, 2023. doi: 10.5281/zenodo.8277473.
[4]  C. R. Harris *et al.*, "Array programming with NumPy," *Nature*, vol. 585, no. 7825, Art. no. 7825, Sep. 2020, doi: 10.1038/s41586-020-2649-2.
[5]  A. Miles *et al.*, "zarr-developers/zarr-python: v2.16.1," *Zenodo*, Aug. 2023, doi: 10.5281/zenodo.8263439.
[6]  B. Kemp, Aeilko Zwinderman, B. Tuk, H. Kamphuisen, and J. Oberyé, "The Sleep-EDF Database [Expanded]." physionet.org, 2018. doi: 10.13026/C2X676.
[7]  A. L. Goldberger *et al.*, "PhysioBank, PhysioToolkit, and PhysioNet: components of a new research resource for complex physiologic signals," *circulation*, vol. 101, no. 23, pp. e215–e220, 2000.
[8]  A. Guillot, F. Sauvet, E. H. During, and V. Thorey, "Dreem Open Datasets: Multi-Scored Sleep Datasets to Compare Human and Automated Sleep Staging," *IEEE Trans. Neural Syst. Rehabil. Eng. Publ. IEEE Eng. Med. Biol. Soc.*, vol. 28, no. 9, pp. 1955–1965, Sep. 2020, doi: 10.1109/TNSRE.2020.3011181.
[9]  H. Nahrstaedt *et al.*, "holgern/pyedflib: v0.1.23," *Zenodo*, doi: 10.5281/zenodo.5678481.
[10] A. Collette *et al.*, "h5py/h5py: 3.8.0," *Zenodo*, doi: 10.5281/zenodo.7568214.
[11] M. Rusanen *et al.*, "aSAGA: Automatic Sleep Analysis with Gray Areas." arXiv, Oct. 03, 2023. doi: 10.48550/arXiv.2310.02032.